\documentstyle[prb,aps]{revtex}
\begin{document}
\draft
\twocolumn[\hsize\textwidth\columnwidth\hsize\csname@twocolumnfalse\endcsname

\title{Microscopic calculations on Raman scattering from acoustic phonons confined
in Si nanocrystals}
\author{Jian Zi\cite{byline1}}
\address{Surface Physics Laboratory, Fudan University, Shanghai 200433, P. R. China}
\author{Kaiming Zhang and Xide Xie}
\address{CCAST (World Laboratory), P. O. Box 8730 Beijing and\\
Surface Physics Laboratory, Fudan University, Shanghai 200433, P. R. China}
\maketitle

\begin{abstract}
Raman scattering from the acoustic phonons confined in Si nanocrystals is
investigated by a microscopic lattice dynamical calculation. Polarized and
depolarized Raman spectra are found to be rather different, indicating from
different acoustic phonons. The polarized scattering is more pronounced than
the depolarized one. The calculated results are compared with the recent
experimental data. The effects of matrix are discussed.
\end{abstract}


] \newpage\narrowtext

Recently, considerable attention has been paid to Si nanostructures since
the discovery of the efficient photoluminescence from porous Si at room
temperature. \cite{can:90} Raman spectroscopy has been intensively used to
characterize porous Si \cite{sui:92,mun:92,tan:96} and Si nanocrystals. \cite
{fur:88,koz:91} A phenomenological phonon-confinement model \cite
{ric:81,cam:86} has been widely used to quantitatively describe the Raman
scattering from optical phonons. Very recently, microscopic calculations
have been carried out to study the Raman spectra from the optical phonons
confined in Si nanocrystals by the authors. \cite{zi:96,zi:97a}

As a matter of fact, when the sizes of nanocrystals decrease, phonons with
large wave-vectors will be involved in the electron-phonon interaction. As a
result, the scattering of electrons by acoustic phonons is expected to be
more pronounced as compared to that by optical phonons. Experimentally,
Raman scattering from confined acoustic phonons has been reported in
glasses, \cite{dua:86} metals \cite{mar:88,fuj:91,fuj:92} and semiconductor
nanocrystals. \cite{ovs:88,cha:93,tan:93,ovs:96} More recently, Raman
spectra from the acoustic phonons confined in Si nanocrystals dispersed in
SiO$_2$ thin films have been reported. \cite{fuj:96}

On the theoretical side, the confined acoustic phonons in nanocrystals have
been studied by assuming a spherical elastic continuum based on the theory
developed by Lamb \cite{lam:82,tam:82} more than a century ago. The
application of this theory to Si nanocrystals is, however, far from
satisfactory. \cite{fuj:96} The first reason may stem from the
oversimplified assumption that the nanocrystal sphere is elastically
isotropic, whereas Si is highly elastically asymmetric. The second is that
the assumption of an elastic continuum may not be valid for nanocrystals
with small size. The predicted Raman frequencies of the confined acoustic
phonons in Si nanocrystals by using Lamb's theory are much higher than the
experimental data. \cite{fuj:96} As pointed out by Fujii {\it et al}. \cite
{fuj:96} that in this case lattice dynamical calculations are desired, which
motivates us to develope a microscopic calculation.

Nanocrystals are normally dispersed in some matrices. The effects of
surrounding matrix on the confined acoustic phonons have recently been
studied by several authors \cite{ovs:96,tam:82,mon:95} using an elastic
continuum theory. Contradictory conclusions were made. Ovsyuk and Novikov 
\cite{ovs:96} claimed that the matrix effects are important, while Montagna
and Dusi \cite{mon:95} reported that the influences of matrices are rather
small and can be negligible. We try to clarify this problem by a lattice
dynamical calculation from a microscopic point of view.

In the present work, we calculate the Raman spectra from the acoustic phonon
confined in Si nanocrystals by a lattice dynamical calculation. The Si
nanocrystals are assumed to have spherical shapes. In our lattice dynamical
calculations, it is assumed that Si atoms are located at their diamond
lattice sites and no relaxation exists as has been done in the previous work
to study the optical phonons. \cite{zi:96,zi:97a} As a first approximation,
the force constants in Si nanocrystals are taken to be the same as those in
the bulk. A partial density approach is adopted to calculate the force
constants in crystalline Si. \cite{fal:88,bus:94} Force constants up to the
fifth nearest neighbors are considered.

The spectral density of the scattered light is given by \cite{ber:76}

\begin{equation}
I_{\mu \nu }({\bf q},\omega )\propto \int dt\exp (-i\omega t)\ \langle
\delta \epsilon _{\mu \nu }^{*}({\bf q},0)\delta \epsilon _{\mu \nu }({\bf q}%
,t)\rangle ,
\end{equation}
where $\mu $ and $\nu $ are the polarization direction of the incident and
scattered photon; $\hbar \omega =\hbar \omega _i-\hbar \omega _s$ and ${\bf q%
}={\bf k}_i-{\bf k}_s$ are the exchanged energy and wave-vector. The
fluctuations of the dielectric constant can be described in terms of the
space Fourier transformation of the macroscopic polarizability density
tensor $P_{\mu \nu }({\bf r},t)$

\begin{eqnarray}
\delta \epsilon _{\mu \nu }({\bf q},t) &\propto &\int d{\bf r}\exp [-i{\bf %
q\cdot r}(t)]P_{\mu \nu }({\bf r},t)  \nonumber \\
&=&\sum_i\exp [-i{\bf q\cdot r}^i(t)]\alpha _{\mu \nu }^i(t).
\end{eqnarray}
Here $\alpha _{\mu \nu }^i(t)$ is the instantaneous polarizability of the $i$%
th scatter at the instantaneous position ${\bf r}^i(t)={\bf R}^i+{\bf u}%
^i(t) $, where ${\bf R}^i$ is the equilibrium position of the $i$th scatter
and ${\bf u}^i$ is the displacement from the equilibrium due to the phonon
vibrations. Since we are interested in Raman scattering from nanocrystals
with size $L$ small compared with the wavelength of the light ($qL\ll 1$),
and not in the Brillouin scattering, we have ${\bf q\sim }$0. The effective
microscopic polarizability $\alpha _{\mu \nu }^i(t)$ can be expanded in
terms of the displacements ${\bf u}^i$, and ${\bf u}^i$ can be expressed in
terms of the vibrational eigenvectors ${\bf e}(i,p)$, the frequency of which
is $\omega _p$. The contribution of the $p$th phonon mode to the Stokes part
of the Raman spectra is then given by \cite{ben:91}

\begin{equation}
I_{\mu \nu }(\omega _p)\propto \frac{n(\omega _p,T)+1}{\omega _p}C_{\mu \nu
}(\omega _p),
\end{equation}
where $n(\omega ,T)$ is the Bose-Einstein population factor at temperature $%
T $ and $C_{\mu \nu }(\omega _p)$ is the mode-radiation coupling coefficient
given by

\begin{equation}
C_{\mu \nu }(\omega _p)=\left| \sum_{ij}\sum_\gamma \frac{\partial \alpha
_{\mu \nu }^i}{\partial u_\gamma ^j}[e_\gamma (j,p)-e_\gamma (i,p)]\right|
^2.
\end{equation}
The quantities $A_{\mu \nu \gamma }^{ij}=\partial \alpha _{\mu \nu
}^i/\partial u_\gamma ^j$ depend on the scattering mechanism. A
bond-polarizability (BP) model \cite{jus:89} is adopted in the present work
to deal with the quantities $A_{\mu \nu \gamma }^{ij}$. The detailed
description of the BP model can be found elsewhere. \cite{zi:97b} Within the
frame work of BP model, the polarizability of the whole system is calculated
as a sum of independent contributions from every bond in the system based on
the calculated eigenvalues and eigenvectors. The quantities $A_{\mu \nu
\gamma }^{ij}$ are nonzero only if $i$ and $j$ are nearest atoms. The Raman
intensity in the $\mu \nu $ polarization for backscattering configuration is
finally given by

\begin{equation}
I_{\mu \nu }(\omega )\propto \sum_p\frac{n(\omega _p,T)+1}{\omega _p}\delta
(\omega -\omega _p)C_{\mu \nu }(\omega _p).
\end{equation}
Neither Fr\"{o}lich interactions nor electro-optic effects are incorporated.
The polarized and depolarized Raman spectra $I_p$ and $I_d$ are obtained by
averaging over the different directions of polarization

\begin{eqnarray}
I_p &=&\frac 13\left( I_{xx}+I_{yy}+I_{zz}\right) , \\
I_d &=&\frac 13\left( I_{xy}+I_{yz}+I_{zx}\right) .
\end{eqnarray}

By using the force constants the dynamical matrix of a Si nanocrystal can be
constructed. Eigenfrequencies and eigenvectors can be obtained by solving
the secular equation about the dynamical matrix. Raman spectra are then
calculated by the BP model based on the obtained eigenfrequencies and
eigenvectors.

Figure \ref{fig1} shows the calculated polarized and depolarized Raman
spectra from the acoustic phonons confined in Si spheres without matrix. For
a Si sphere the size is measured by its diameter, given by $L=(3N/4\pi
)^{1/3}a$, where $N$ and $a$ are the number of Si atoms and the lattice
constant of crystalline Si, respectively. Raman peaks can be clearly seen in
each spectrum. In the polarized spectrum, there is a Raman peak, labeled as
peak 2. By inspecting the atomic displacements, LA-like phonons are found to
be responsible for this peak. There is a small peak (peak 4) at the higher
frequency, which originates from the higher-order LA-like confined phonons
similar to the higher-order folded LA phonons in superlattices. \cite{zi:97b}
In depolarized spectrum, there is a Raman peak, labeled as peak 1. TA-like
phonons are responsible for this peak. A rather small peak (peak 3) with the
same frequency as the polarized peak 2 exists, which indicates that LA-like
phonons also contribute to the depolarized scattering. As the size
decreases, the peaks 1-4 shift to higher frequency. The intensity of the
depolarized peak 1 is always smaller than that of the polarized peak 2. The
intensity ratio of the two peaks in the depolarized and polarized scattering
is roughly about 0.3 for all Si spheres studied here, which agrees fairly
well with the experimental result of 0.25. \cite{fuj:96} From Fig. \ref{fig1}%
, it can be seen that the peak positions depend strongly on the sizes of Si
spheres.

In order to get the insight into the size dependencies, the peak frequencies
as a function of the inverse size are given in Fig. \ref{fig2}. The bulk
bands of sound waves, obtained by a simple correspondence between the
wave-vector and size $q=\pi /L$, are also given as hatched areas for
reference. This correspondence should be valid at least for very large size.
For sound waves propagating in crystalline Si, only in some high-symmetrical
directions are these waves purely longitudinal or transverse. In general,
they have components of both. In an arbitrary direction quasi-longitudinal
(QL) and quasi-transverse (QT) modes are obtained. The upper hatched area
corresponds to the QL bands and the lower one to the QT bands.

The confined acoustic phonons in an isotropically elastic sphere were
previously studied by Lamb's theory. \cite{lam:82,tam:82} Two types of
confined acoustic modes, spheroidal and torsional modes, were derived. The
frequencies of these two modes were found to be proportional to the sound
velocities in spheres and inversely proportional to the sphere size. The
spheroidal and torsional modes are characterized by a quantum number $l$.
From the selection rules, Raman-active modes are spheroidal modes with $l=0$
and 2. \cite{duv:92} The $l=0$ mode produces totally polarized spectra,
while the $l=2$ one partially depolarized spectra. This model has been used
in many previous studies and could explain some experimental results. The
results based on Lamb's theory are also given in Fig. 2 as dashed lines. The
lines for every $l$ stand for different propagating directions, since the
sound velocity of Si is different for different propagating direction.

It can be seen from Fig. \ref{fig2} that the results from Lamb's theory are
systematically two times as larger as the experimental data. It should be
noted that in experiment Si nanocrystals are surrounded by SiO$_2$ matrix.
Even by taking the matrix effects into account, the results predicted by
Lamb's theory are still much larger than the experimental ones.

The results obtained by the lattice dynamical calculation given in the
figure are for Si nanocrystals without any matrix. The frequencies of the
polarized peaks are outside the QL bands, which indicates that the effective
sound velocity for LA-like phonons is lower than the bulk counterpart and is
somewhat {\it softened }due to the finite size. The frequencies of the
depolarized peaks are just inside the QT bands for Si nanocrystals without
matrix. It seems that the polarized and depolarized peak frequencies scale
almost linearly with the inverse of size in the size range studied, as is
predicted by Lamb's theory. By a careful analysis, we find that this is,
however, not the case owing to the fact that the Raman peak frequencies
should gradually merge into the bulk bands with the increase in size.

We do not attempt to compare our calculated results directly with the
experimental ones, since the calculated results are without matrix, while Si
nanocrystals in experiment are surrounded with SiO$_2$ matrix. In principle,
we can study Si nanocrystals with any matrix and compare directly with
experiment. There are, however, some difficulties: there is little
information on the structure of matrix, the size of matrix, and the
situation about the interface between nanocrystals and matrices. Our
calculated results of Si nanocrystals without matrix are larger than the
experimental data. We will show below that this disagreement is caused by
the matrix effects. It is expected that the results of the lattice dynamical
calculation should agree better with the experimental data for Si
nanocrystals with large size since the matrix effects should be less
important for nanocrystals with large size as. This can be seen by a simple
extrapolation of the data obtained by a lattice dynamical calculation in the
direction towards the larger size.

As mentioned above, there are controversies about the effects of matrices.
To clarify this problem, the effects of matrix are studied by introducing
some shells of matrix atoms. We must simplified the problem since we have
little about the matrix as mentioned above. The matrix atoms are assumed to
be located still at the diamond lattice and the force constants of the
matrix atoms are the same as in those of Si nanocrystals. The only
difference is that the matrix atom has an {\it effective }mass $M_x$. The
calculated results are shown in Fig. \ref{fig3}. Our calculations reveal
that the frequencies of the polarized and depolarized Raman peaks are very
sensitive to the matrix. The frequencies of both polarized and depolarized
Raman peaks are found to shift to lower frequencies owing to the effects of
matrix. The downward frequency shift due to the matrix can explain the
discrepancy between our results without matrix and the experimental ones.

Other set of force constants between matrix atoms is also probed. The
conclusion is basically the same. The difference is a different amount of
frequency shift. The importance of the effects of matrix on acoustic phonons
can be understood by that fact that the vibrational amplitudes of acoustic
phonons are large at the boundary. For optical phonons, the vibrational
amplitudes are very small at the boundary that the Raman frequencies should
be slightly affected by the matrix. \cite{zi:97a} This has been demonstrated
by our calculations.

In summary, we have investigated the Raman scattering from acoustic phonons
confined in Si nanocrystals by a lattice dynamical calculation. The
polarized and depolarized low-frequency Raman peaks originate from the
confined LA-like and TA-like acoustic phonons, respectively. The polarized
scattering is more pronounced than the depolarized one. The effects of
matrix are important and will lead to a downward frequency shift for both
polarized and depolarized Raman peaks.

This work is supported in part by the NSF of China under Contract No.
69625609 and by National PAN-DENG project (Grant No. 95-YU-41).

\begin{figure}[tbp]
\caption{Calculated Raman spectrea from the acoustic phonons confined in Si
spheres with different sizes at room temperature. The solid (dashed) lines
stand for polarized (depolarized) Raman spectra.}
\label{fig1}
\end{figure}

\begin{figure}[tbp]
\caption{Raman peak frequencies from acoustic phonons confined in Si
nanocrystals versus the inverse of size. The results obtained by the lattice
dynamical calculation are given as empty (solid) circles for polarized
(depolarized ) scattering for nanocrystals without matrix. The hatched areas
are the bulk bands. The experimental results, taken from Ref. 
\protect\onlinecite{fuj:96}, are also given as empty (solid) squares for polarized
(depolarized) peaks, respectively. Dashed lines are results by Lamb's
theory. }
\label{fig2}
\end{figure}

\begin{figure}[tbp]
\caption{Calculated Raman frequency shifts $\Delta \omega =\omega -\omega _0$
for a Si nanocrystal of 2.39 nm owing to a matrix with a width of 0.39 nm,
where $\omega $ and $\omega _o$ are the frequecny with and without a matrix.}
\label{fig3}
\end{figure}

\end{document}